\documentclass[prx,twocolumn]{revtex4}

\usepackage{amsmath,amssymb} 
\usepackage{bm}
\usepackage[colorlinks=true, linkcolor=blue, urlcolor=blue, citecolor=blue]{hyperref}
\usepackage{subfigure}
\usepackage{graphicx}
\usepackage{color}
\usepackage[dvipsnames]{xcolor}
\usepackage{subfigure}
\usepackage{sidecap}
\usepackage{pgfplots}
\usepackage{tikz}
\usepackage{cleveref}
\usepackage{lettrine}
\write18{-shell-escape}
\usetikzlibrary{external}
\usetikzlibrary{datavisualization}
\usetikzlibrary{matrix}
\usepackage{pgfplots}
	\usepgfplotslibrary{groupplots}
\pgfplotsset{width = 1in, compat=1.3}

\renewcommand{\vec}[1]{{\mathbf #1}}
\newcommand{\unitv}[1]{\hat{\bf #1}}

\newcommand{\paren}[1]{\left( #1 \right)}
\newcommand{\ket}[1]{|\rule{0cm}{2ex} #1 \rangle}

\newcommand{\xhat}{\unitv{x}}    
\newcommand{\yhat}{\unitv{y}}
\newcommand{\zhat}{\unitv{z}}

\newcommand{\kplus}{\ket{\!+\!v_B}}
\newcommand{\kminus}{\ket{\!-\!v_B}}

\colorlet{Col1}{blue}
\colorlet{Col2}{red}

\colorlet{Col3}{ForestGreen}
\colorlet{Col4}{orange}

\begin{document}

\title{Quantum Rotation Sensing with Dual Sagnac Interferometers\\
in an Atom-Optical Waveguide}

\date{\today}

\author{E. R. Moan}
\author{R. A. Horne}
\altaffiliation[Present address: ]{NASA Langley Research Center, Revolutionary Aviation Technologies Branch, Mail Stop 207, Hampton, Virginia 23681, USA}
\author{T. Arpornthip}
\altaffiliation[Present address: ]{Prince of Songhkla University, Interdisciplinary Graduate School of Earth System Science and Andaman Natural Disaster Management, Kathu, Phuket 83120, Thailand}
\author{Z. Luo}
\author{A. J. Fallon}
\author{S. J. Berl}
\author{C. A. Sackett}
\email[Email: ]{cas8m@virginia.edu}
\affiliation{Department of Physics, University of Virginia, Charlottesville, Virginia 22904, USA}

\begin{abstract}
We describe a Sagnac interferometer suitable for rotation sensing, implemented using an atomic Bose-Einstein condensate confined in a harmonic magnetic trap.  The atom wave packets are split and recombined
by standing-wave Bragg lasers, and the trapping potential steers the packets along circular trajectories with a radius of 0.2~mm. Two conjugate interferometers are implemented simultaneously to provide common-mode rejection of noise and to isolate the rotation signal.  With interference visibilities of about 50\%, we achieve a rotation sensitivity comparable to Earth's rate in about 10~min of operation. Gyroscope operation was demonstrated by rotating the optical table on which the experiment was performed.
\end{abstract}

\maketitle

Sensitive and accurate rotation sensing is a critical requirement for applications such as inertial navigation \cite{Grewal13}, north-finding \cite{Prikhodko13}, geophysical analysis \cite{Stedman97}, and tests of general relativity \cite{Cerdonio88}. One effective technique used for rotation sensing is Sagnac interferometry, in which a wave is split, traverses two paths that enclose an area, and then recombined. The resulting interference signal depends on the rotation rate of the system and the area enclosed by the paths \cite{Sagnac1913}. Optical Sagnac interferometers are an important component in present-day navigation systems \cite{Lefevre14}, but suffer from limited sensitivity and stability. Interferometers using matter waves are intrinsically more sensitive and have demonstrated superior gyroscope performance \cite{Gustavson00,Durfee06,Savoie18}, but the benefits have not been large enough to offset the substantial increase in apparatus size and complexity that atomic systems require. It has been suggested that these problems might be overcome using atoms confined in a guiding potential or trap, as opposed to atoms falling in free space \cite{Ketterle92,Barrett01,Arnold04}.  The trap can support the atoms against gravity, so a long measurement time can be achieved without requiring a large drop distance. The trap can also control the trajectory of the atoms, causing them to move in a circular loop that provides a large enclosed area for a given linear size \cite{Horikoshi07b}.  Here we use such an approach to demonstrate a rotation measurement with Earth-rate sensitivity.

A small number of trapped-atom Sagnac interferometers have been demonstrated in the past \cite{Jo07,Wu07,Burke09,Qi17,Boshier19}, but none have been used to make a quantitative rotation measurement. The largest enclosed areas have been achieved using a linear interferometer that is translated along a direction perpendicular to the interferometer axis \cite{Wu07,Boshier19}, but this approach may not be wellsuited for inertial measurements in a moving vehicle. Here, we demonstrate a true two-dimensional interferometer configuration in which atoms travel in circular trajectories through a static confining potential. We obtain an effective enclosed area of $0.50$~mm$^2$, compared to areas of $0.20$~mm$^2$ reported by Wu~{\em et al.} \cite{Wu07} and $0.35$~mm$^2$ recently obtained by the Boshier~{\em et al.} \cite{Boshier19}.  We argue below that our effective area can be scaled up to 1~cm$^2$ or more, which would offer sensitivity sufficient for many practical applications.

Another key advance is the use of dual counterpropagating interferometer measurements. Here, two Sagnac interferometers are implemented at the same time in the same trap, using atoms traveling with opposite velocities over the same paths. This technique was developed for free-space interferometers \cite{Durfee06} and allows common-mode rejection of phase noise that can otherwise mask the rotation signal. The Sagnac effect itself is differential and can be extracted by comparing the two individual measurements. This technique is likely to be essential for a practical rotation-sensing system, but it has not previously been demonstrated in a trapped-atom device.

Our Sagnac interferometer is implemented using a Bose-Einstein condensate confined in a three-dimensional trap, with potential energy
\begin{equation}
V(x,y,z) = \frac{1}{2}m\paren{\omega_x^2 x^2 + \omega_y^2 y^2 + \omega_z^2 z^2},
\end{equation}
where $m$ is the atomic mass and the $\omega_i$ are the trap frequencies. We consider first the ideal case where the trap is
cylindrically symmetric $\omega_x = \omega_y = \omega_0$, and the atoms form a pure condensate with negligible interactions at
rest in the center of the trap. 
The atoms are manipulated using a set of standing-wave Bragg lasers \cite{Wu05b,Hughes07} propagating along the $x$ and $y$ directions.  The beams couple atomic states with momenta $\vec{p}$ and $\vec{p}\pm 2\hbar\vec{k}$, where $\vec{k}$ is the wave vector of the laser. We express this in terms of a Bragg velocity kick $v_B = 2\hbar k/m$.  

The interferometer measurement begins by applying the Bragg beams along $y$. This generates two wave packets with velocities $\vec{v} = \pm v_B \unitv{y}$. The wave packets move in the trap, with their centers of mass following the ordinary trajectory for a harmonic oscillator, $x(t) = 0$ and $y(t) = \pm(v_B/\omega_0)\sin\omega_0 t$. After a time $t_1 = \pi/(2\omega_0)$, the atoms come to rest near the classical turning point at a radius $R = v_B/\omega_0$. The Bragg beams traveling along $x$ are then applied to both packets, providing velocity kicks $\pm v_B \unitv{x}$ and generating a total of four packets. Each of these packets now travels in a circle with radius $R$, as $x(t) = \pm R\sin\omega_0 t$ and $y(t) = \pm R\cos\omega_0 t$. The atoms propagate for time $t_2 = 2\pi/\omega_0$, completing one full orbit around the trap. Figure~1(a) illustrates the packet trajectories.

After the orbit, the pairs of packets are overlapped again at their locations prior to the $x$ Bragg pulse. The wave function of a pair can be expressed as
\begin{equation}
\ket{\psi} = \frac{1}{\sqrt{2}} \paren{e^{i\Phi/2} \kplus + e^{-i\Phi/2} \kminus},
\end{equation}
where $\Phi$ is the phase developed between the packets and $\ket{\!\pm\! v_B}$ are states with the indicated velocity along $x$.  The $x$ Bragg beams are now applied again. The even superposition $(\kplus + \kminus)$ is coupled back to the zero-momentum state $\ket{0}$, while the odd superposition $(\kplus - \kminus)$ remains unchanged (up to an overall phase). Projecting the wave function $\ket{\psi}$ onto this basis, we find that a fraction $S = \cos^2 (\Phi/2)$ of the atoms are brought back to rest \cite{Wang05,Garcia06}.

To detect the result of the recombination, the moving and stationary atoms are allowed to separate, the trap is turned off, and the entire system is observed using absorption imaging in the $xy$-plane. Figure~1(b) shows the measured positions of one packet as it traverses the trap. A video showing the motion of all four packets is available in the Supplemental Material \cite{Moan_supp}.  Figure~2(a) shows a typical absorption image after the recombination pulse.

\begin{figure}
	\includegraphics{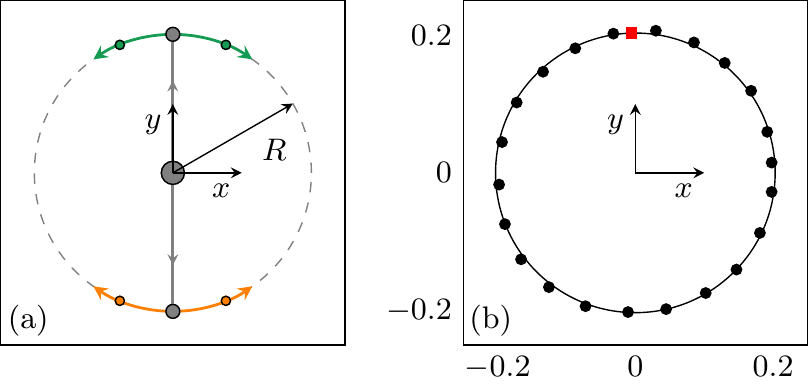}
	\caption{(a) Trajectories of atoms in the interferometer.  The initial condensate (center) is split into two packets that move along $\pm y$. When the packets reach their turning points at $y = \pm R$, they are split along $x$, generating four packets that move in circular orbits (green and orange traces). After one complete orbit the packets are recombined, forming two independent interferometers. (b) Experimental data showing the path of one packet, starting at the red square and moving counterclockwise in 5~ms increments. The scales are in mm.  	  
	}
\end{figure}

This sequence produces two independent interferometer measurements. We define $\Phi_+$ as the phase measured for atoms at $y = +R$, and $\Phi_-$ as the phase measured at $y = -R$. Each of these individual phases is sensitive to a variety of effects including magnetic field variations, mechanical vibration, and laser phase noise.  Most of these effects will be the same for both interferometers, but if the system is rotating with angular velocity $\boldsymbol{\Omega}$, then the Sagnac phases
\begin{equation}
\Phi_S = \frac{1}{\hbar} \oint \Delta\vec{L}\cdot\boldsymbol{\Omega}\,dt = \frac{4 m\Omega A}{\hbar}
\end{equation}
will have the opposite sign. Here $\Delta\vec{L} = \vec{r}_2\times\vec{p}_2 - \vec{r}_1\times\vec{p}_1$ is the difference in angular momentum between the two packets, and $A = \pi R^2$ is the area of a single packet's orbit.  This leads to a differential phase
\begin{equation}
\Delta \Phi = \Phi_+ - \Phi_- =  2\Phi_S = \frac{8 m \Omega A}{\hbar}.
\label{eq: 9}
\end{equation}

We implement the interferometer using approximately $10^4$ $^{87}$Rb atoms that are magnetically trapped in the $F = 2, m_F =2$ ground state.  The apparatus for condensate production and generating the magnetic trap has been previously described \cite{Horne17}. Our time-orbiting potential (TOP) trap \cite{Petrich95} uses a special field configuration that allows precise experimental control of the trap parameters.  The bias field of the TOP trap rotates as
\begin{align}\label{eq:Bias}
	\vec{B}_{\rm b}& = B_0\Big[ (1+\alpha)\sin (\Omega_1 t) \cos (\Omega_2 t + \beta) \, \xhat \nonumber \\
		& + (1-\alpha) \sin(\Omega_1 t) \sin (\Omega_2 t - \beta) \, \yhat + \cos (\Omega_1 t) \, \zhat \Big],
\end{align} 
with $\Omega_1 \approx 2\pi\times 10^4$~Hz and $\Omega_2 \approx 2\pi\times 10^3$~Hz. The amplitude asymmetry $\alpha$ and the phase $\beta$ are nominally zero, but can be adjusted to optimize the trap. The coordinates $(x,y,z)$ are defined by the coil geometry, with $z$ near vertical. The trap also uses an oscillating quadrupole field,
\begin{equation}
	\vec{B}_{\rm q} = \frac 1 2 B_1' \cos (\Omega_1 t) \big(x \, \xhat + y \, \yhat - 2 z \, \zhat \big).
\end{equation}
These fields provide a trap potential $V(\vec{r}) = \mu_B \langle|\vec{B}|\rangle$, where $\mu_B$ is the Bohr magneton and $\langle|\vec{B}|\rangle$ is the time average of the magnetic field magnitude. Evaluation to second order in the coordinates gives
\begin{multline}  \label{eqV}
V = -\frac{1}{2}\mu_B B_1 z + \frac{1}{2}m\omega_0^2  \Bigg[
\left( 1+\frac{2\alpha}{7} \right) x^2 \\ 
 + \left(1-\frac{2\alpha}{7} \right) y^2  
 +\, \frac{4}{7} \beta xy +\frac{8}{7}z^2 \Bigg],
\end{multline}
with $\omega_0 = (7\mu_B B_1'^2/64mB_0)^{1/2}$. We set $B_1' = 2mg/\mu_B \approx 31$~G/cm to cancel gravity at the center of the trap, and with $B_0 \approx 2$~G we obtain $\omega_0 \approx 2\pi\times 9$~Hz.

The actual potential experienced by the atoms differs from Eq.~\eqref{eqV}.
For instance, we measure the oscillation frequency along $z$ to be about 11~Hz, 
which differs from the 9.6~Hz prediction due to a combination of curvature in the trap bias
fields and unbalanced bias amplitudes in the horizontal and vertical directions.
If effects such as these alter the wave packet trajectories so that they fail to overlap in both position and momentum after an orbit,  then the interferometer will be spoiled. A classical trajectory calculation indicates that the difference in trap frequencies $|\omega_x-\omega_y|/2\pi$ must be less than 0.1 Hz, and the Bragg laser beams must be aligned to better than 10~mrad accuracy. We achieve these requirements by observing the packet trajectories and adjusting $\alpha$, $\beta$, the Bragg beam angles, and the propagation times $t_1$ and $t_2$ until both interferometers are closed and interference is observed. The resulting orbits are nearly circular with a radius $R = 0.20$~mm, as shown in Fig.~1. The packets can also be imaged in the $yz$ plane, which is used to ensure overlap in the $z$ direction.

\begin{figure}
    \includegraphics{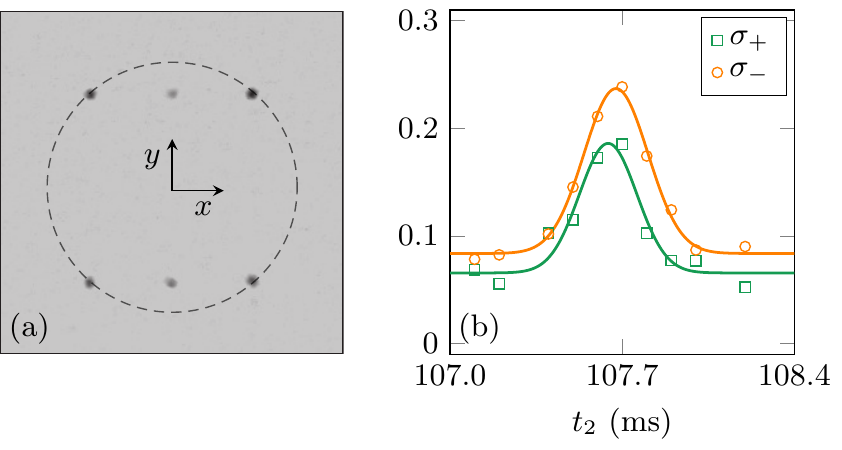}
	\caption{(a) An absorption image in the $xy$ plane taken 12~ms after the recombination pulse was applied.  Atoms that were brought back to rest now oscillate along $y$, while the remaining atoms continue to move along the dashed circle. Here, the fraction of atoms brought to rest is small, indicating that both interferometers measure a phase close to $\pi$. (b) The standard deviations $\sigma_+$ and $\sigma_-$ of the two interferometer output signals, $S_+$ and $S_-$, are plotted as the interferometer time $t_2$ is varied. When the interference contrast is high, $\sigma$ is large because of phase noise from vibrations and other sources. 	At $t_2 = 107.7$~ms, both interferometers exhibit good contrast.  
	\label{fig: fig2}
	}
\end{figure}

When the Bragg beam and trap parameters are optimized, we observe simultaneous interference for both packet pairs. Each individual interferometer exhibits phase noise, primarily due to vibrations of the mirror used to retro-reflect the $x$ Bragg beam. The appearance of this noise is an indicator that interference is occurring, as shown in Fig.~2(b).  The duration of the contrast peaks indicates a coherence length of
10~$\mu$m, which agrees with the Thomas-Fermi condensate size for our trap \cite{Dalfovo99}. 
The noise peak is consistent with an interference visibility of about 50\%, which could be attributed to
residual alignment errors.

In reality, the condensate source is not at zero temperature. We generally observe no
noncondensate atoms, but given our imaging resolution the condensate fraction could still be as low as
60\%. This corresponds to a temperature of about 7~nK, which is cold enough for the thermal atoms to be split efficiently by the Bragg lasers \cite{Hughes07}. We do not observe interference
for noncondensed atoms in the trap, so the presence of a thermal fraction could also contribute to the reduced visibility. 

The lifetime of the condensate is about 60~s, which is much longer than our 0.11~s
interferometer time. The peak density of a single packet is about $3\times 10^{12}$~cm$^{-3}$, 
leading to a 2\% chance for an atom to be lost via elastic collisions as two packets pass through each other. 
The chemical potential of the initial condensate is about $2\pi\hbar\times 40$~Hz,
so no internal condensate dynamics can occur during the 0.5-ms time that two wave packets interact. The Bragg beam can excite breathing motion of the packets due to the rapid decrease in interaction energy after splitting. We have numerically simulated these excitations and find that they have a negligible effect on the packet's phase evolution \cite{Fallon15}. Any overall phases induced by interactions will be common mode and cancel in $\Delta\Phi$.

Each interferometer's output signal is $S = N_0/N$, the fraction of atoms brought back to rest.  When the two signals $S_+$ and $S_-$ are plotted against each other, the data fall on an ellipse, with the eccentricity and orientation of the ellipse set by  the differential phase \cite{Foster02}. Example data are shown on the left in Fig.~3.  We fit such data to an ellipse to extract the phase $\Delta\Phi$, with an accuracy of about $0.2$ rad after ten runs of the experiment.

\begin{figure}
	\includegraphics{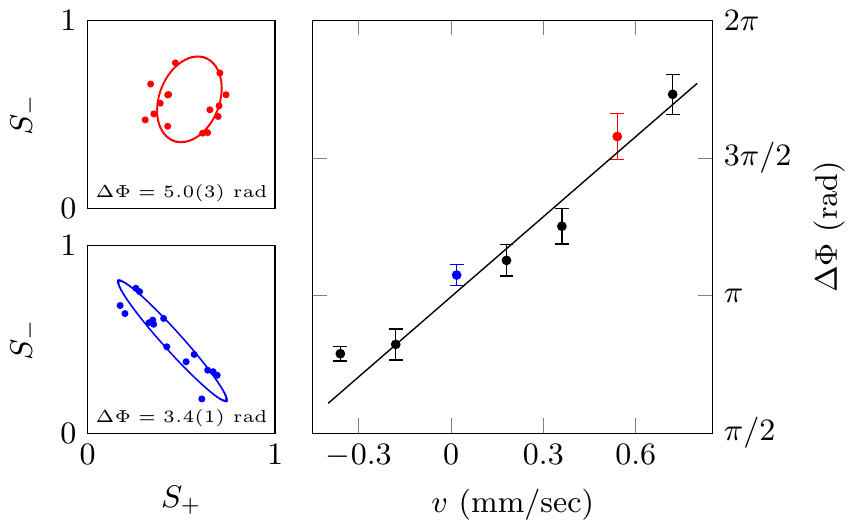}
	\caption{Left: Points correspond to the two interferometer output signals $S_+$ and $S_-$ from a given measurement. Curves are ellipses fitted to the points. The orientation and eccentricity of the ellipse reveals the differential phase $\Delta\Phi$ between the interferometers. Right: Dependence of the differential phase on the rotation velocity of the optical table $v$ on which the experiment rests, illustrating the Sagnac effect.  The shaded points correspond to the matching data on the left. The nonzero phase  at $v = 0$~mm/s is consistent with a weak anharmonicity of the trap potential.}
	\label{fig: fig3}	
\end{figure}

The effective Sagnac area of the interferometer is $4A = 0.50$~mm$^2$, which implies a
Sagnac phase of about 1.4 rad/(mrad/s). 
In order to observe the Sagnac sensitivity of the interferometer, we rotated the optical table on which the apparatus sits. The table floats on air legs that allow a few mm of horizontal motion.  We used a motorized translation stage to move one end of the table during the interferometer measurement. With a force of about 10~N, linear speeds up to 0.7 mm/s could be achieved. The table motion was initiated prior to the first Bragg pulse of the interferometer sequence, and it continued at a constant speed throughout the measurement. Results are shown in Fig.~3. 

Because the rotation measurements were not made using a dedicated rate table, accurate calibration of the rotation rate is difficult. Applying the Sagnac formula to the data in Fig.~3 indicates a rotation radius $v/\Omega$ of 0.5~m, in reasonable 
agreement with a mechanical estimate of about 1~m.  
The table did not remain level during the rotation, but we experimentally verified that
the phase is insensitive to static tilts, and numerical simulations indicate no significant
sensitivity to dynamic tilting.
The one-sigma error bars in Fig.~3 correspond to a rotation sensitivity of $8\times 10^{-5}$~rad/s, comparable to the rotation rate of the Earth, $\Omega_E = 7.3 \times 10^{-5}$~rad/s. 

This rotation sensitivity is not exceptional, being about what a careful observer might obtain by watching the shadow of a sundial. An attractive sensitivity for inertial navigation applications is $10^{-7}$~(rad/s)/$\sqrt{\mbox{Hz}}$, corresponding to an angle random walk of $3\times 10^{-4}$ deg/$\sqrt{\text{hr}}$ \cite{Grewal13}. We believe this performance level can be reached through the use of a larger enclosed area, more rapid condensate production, and reduction of technical noise. Lowering the trap confinement frequency to 2 Hz would give an orbit radius of 1~mm. To increase the area further, the packets can be allowed to make multiple orbits through the trap. With ten orbits, the total effective area would be 1.3~cm$^2$ with a measurement time of 5~s. An atom-chip apparatus can produce a suitable condensate in less than 5~s with reasonable size, weight and power constraints \cite{Farkas10}. Operating such a device at shot-noise limited phase accuracy would provide the desired sensitivity for inertial navigation.

Implementing these improvements will require careful control of the trapping
potential to avoid extraneous phase noise. The phase difference induced by the trap can be
calculated as \cite{Berman97}
\begin{equation}
    \Phi_{\text{trap}} = \frac{1}{\hbar} \int_{t_i}^{t_f} (\mathcal{L}_1-\mathcal{L}_2)\,dt 
    - \vec{k}_0\cdot \Delta \vec{r}_f,
\end{equation}
where $\mathcal{L}_i = T_i-V_i$ is the Lagrangian evaluated along the classical trajectory of 
packet $i$, $2\hbar\vec{k}_0 = m(\vec{v}_{1f}-\vec{v}_{2f})$ is the difference between
the final momenta of the interfering packets, and 
$\Delta\vec{r} = \vec{r}_{1f}-\vec{r}_{2f}$
is the final packet separation. In an ideal trap these contributions are zero, 
but the phase can be significant in the presence of experimental imperfections.
We have evaluated the differential phase $\Delta\Phi$ 
numerically, under a wide variety of conditions. We find
the dominant contribution from the trap to be
\begin{equation}
   \Delta\Phi_{\text{trap}} \approx -40 kR^3 n(n+1) \gamma c,
\end{equation}
where $k$ is the Bragg laser wave number, $R$ is the orbit
radius, $n$ is the number of orbits, 
and the trapping potential includes nonideal terms 
\begin{equation}
    \delta V = m\omega_0^2\left(\gamma xy + \frac{1}{4}c\rho^4\right).
\end{equation}
The $\gamma$ parameter can be precisely adjusted using the TOP phase $\beta$, 
as in Eq.~\eqref{eq:Bias}. 
The anharmonic term $c$ depends on the coil geometry, and can be adjusted through
the use of additional shim coils. 
By measuring how the oscillation frequency varies with amplitude, we find
that $c \approx -0.3$~mm$^{-2}$ in our trap. For a TOP phase of $\beta = 10^{-2}$~rad, this corresponds to $\delta\Phi_{\text{trap}} \approx 3$ rad, which is consistent with the zero-rotation offset
seen in Fig.~\ref{fig: fig3}. In order to achieve shot-noise-limited
phase stability of $10^{-2}$~rad in the proposed 2~Hz trap, it will be necessary 
to have $|\gamma| < 10^{-5}$ and $|c| < 10^{-4}$~mm$^{-2}$. These are challenging
specifications, but they should be reachable with careful design.

In summary, we have implemented a trapped-atom Sagnac sensor with the largest enclosed area to date, which for the first time uses simultaneous counter-rotating interferometers for common-mode noise rejection and demonstrates actual rotation sensing. The rotation sensitivity is comparable to Earth's rate, and we argue that substantial improvements are feasible. We also note that our interferometer scheme could be of interest for fundamental physics. For example, it could be used to investigate the phase evolution of the trapped atoms themselves, which can exhibit nontrivial behavior such as phase diffusion \cite{Fallon15} and squeezing \cite{Jo07}.

\begin{acknowledgments}
This work has been supported by the National Science Foundation (Grant No. PHY-1607571), NASA (Contract No. RSA1549080), and DARPA (Award No. FA9453-19-1-0007). 
\end{acknowledgments}

\bibliographystyle{apsrev}
\bibliography{Moan.bib}









\end{document}